\documentclass[twocolumn,aps,prl,floatfix,showpacs]{revtex4}
\usepackage{epsfig}
\usepackage{dcolumn}
\usepackage{bm}
\begin{document}

\title{Nonexponential Relaxations in a Two-Dimensional Electron System in Silicon}
\author{J.\ Jaroszy\'nski}
\email{jaroszy@magnet.fsu.edu}
\affiliation{National High Magnetic Field Laboratory, Florida State
University, Tallahassee, Florida 32310}
\author{Dragana  Popovi\'c}
\email{dragana@magnet.fsu.edu} \affiliation{National High Magnetic Field Laboratory, Florida State University,
Tallahassee, Florida 32310}
\date{\today}

\begin{abstract}
The relaxations of conductivity have been studied in a strongly disordered two-dimensional (2D) electron system
in Si after excitation far from equilibrium by a rapid change of carrier density $n_s$ at low temperatures $T$.
The dramatic and precise dependence of the relaxations on $n_s$ and $T$ strongly suggests (a) the transition to
a glassy phase as $T\rightarrow0$, and (b) the Coulomb interactions between 2D electrons play a dominant role in
the observed out-of-equilibrium dynamics.
\end{abstract}


\maketitle

The interplay of strong electronic correlations and disorder is believed to be responsible for a plethora of new
phenomena occurring in many complex materials in the region of the metal-insulator transition
(MIT)~\cite{Mir-Dob-review}.  Even though glassy behavior of electrons is emerging as one of the key concepts,
electron or Coulomb glasses~\cite{eglass} in general remain largely unexplored.  Slow, nonexponential
relaxations characteristic of glassy dynamics have been observed in doped semiconductors~\cite{earlyg}, strongly
disordered InO$_x$ films~\cite{films2,films2-Zvi,films2-Zvi2}, and various granular metals~\cite{films3}, but
the regime near the MIT has been even less studied.  In many complex systems, the MIT is accompanied by changes
in magnetic or structural symmetry, which complicates the situation even further. On the other hand, low-density
two-dimensional (2D) electron and hole systems in semiconductor heterostructures, where the MIT has been a
subject of great interest and debate~\cite{2DMIT-reviews}, represent particularly appealing model systems for
studying the effects of interactions and disorder in a controlled and systematic way.

Our recent resistance noise measurements on a 2D electron system (2DES) in Si~\cite{SBPRL,JJPRL,JJPRLfield} have
shown signatures of glassy behavior at densities $n_s$ lower than some well-defined density $n_g$, such that
$n_g>n_c$ ($n_c$ -- the critical density for the MIT). In particular, by reducing $n_s$, it was observed that
the dynamics suddenly and dramatically slowed down at $n_g$, and there was an abrupt change to a correlated
statistics, consistent with the hierarchical picture of glassy dynamics. These features were shown to persist
even when the 2DES is fully spin polarized~\cite{JJPRLfield}, indicating that the charge degrees of freedom are
responsible for the anomalous noise behavior. Similar results were obtained in both highly disordered
samples~\cite{SBPRL} and those with relatively low disorder~\cite{JJPRL,JJPRLfield}, with $n_g\gtrsim n_c$ in
the latter case.  In addition, for $n_s<n_g$, the conductivity $\sigma$ of the 2DES was found to depend on the
cooling procedure~\cite{SBPRL}, another characteristic of glassy systems. We note that, even though $n_g$ is in
the metallic regime (\textit{i.~e.} $\sigma(T\rightarrow 0)\neq 0$), $\sigma$ is so small that $k_{F}l<1$
($k_{F}$ -- Fermi wave vector, $l$ -- mean free path)~\cite{kfl}, which violates the Mott limit for the metallic
transport in 2D.  Hence the noise studies~\cite{SBPRL,JJPRL,JJPRLfield} as well as this work probe the regime of
strong disorder that encompasses both the insulating phase and the unconventional conducting regime where
$k_{F}l<1$. Such ``bad metals'' include a variety of strongly correlated materials with unusual
properties~\cite{bad}.

Here we report a systematic study of the relaxations (\textit{i.e.} time dependence) of $\sigma$ in the same
highly disordered 2DES in Si after applying a large perturbation at different $n_s$ and $T$.  The perturbation
consists of a large, rapid change of $n_s$, controlled by the gate voltage $V_g$.  For $n_s<n_g$, we find that,
on short enough time scales, $\sigma(t)$ obeys a nonexponential, Ogielski form~\cite{Ogielski}, found in some
other systems~\cite{Pappas} just above the glass transition. The associated relaxation time $\tau_{low}$
diverges as $T\rightarrow 0$, while its density dependence provides strong evidence that the observed
out-of-equilibrium behavior is dominated by the Coulomb interactions between 2D electrons. After a sufficiently
long time $t\gg \tau_{high}$, $\sigma$ relaxes exponentially to its equilibrium value.  The corresponding
characteristic time $\tau_{high}\rightarrow \infty$ as $T\rightarrow 0$, suggesting that the glass transition
temperature $T_{g}=0$. The most peculiar feature of the relaxation is that the system equilibrates only after it
first goes \textit{farther away} from equilibrium.

Measurements were carried out on a 2DES in Si metal-oxide-semiconductor field-effect transistors (MOSFETs) with
a large amount of disorder (the peak mobility $\approx$~0.06~m$^2$/Vs at 4.2~K with the applied back-gate bias
of $-2$~V).  The sample dimensions $L\times W$ ($L$-- length, $W$-- width) were $1\times 90~\mu$m$^2$ and
$2\times 50~\mu$m$^2$. Both samples exhibited the same behavior and, in general, they were almost identical to
the samples used previously in noise measurements~\cite{SBPRL}. The data obtained on a $2\times 50~\mu$m$^2$
sample are presented in detail below. For this sample, $n_s(10^{11}$cm$^{-2})=4.31(V_g[$V$]-6.3)$;
$n_g(10^{11}$cm$^{-2})\approx 7.5$ and $n_c(10^{11}$cm$^{-2})\approx 4.5$, where $n_g$ was determined from noise
and $n_c$ from $\sigma(n_s,T)$ measurements on both metallic and insulating sides (see
Refs.~\onlinecite{SBPRL,JJPRLfield}). The samples and the standard ac lock-in technique (typically 13 Hz) that
was used to measure $\sigma$ have been described in more detail elsewhere~\cite{SBPRL}.

\begin{figure}
\centerline{\epsfig{file=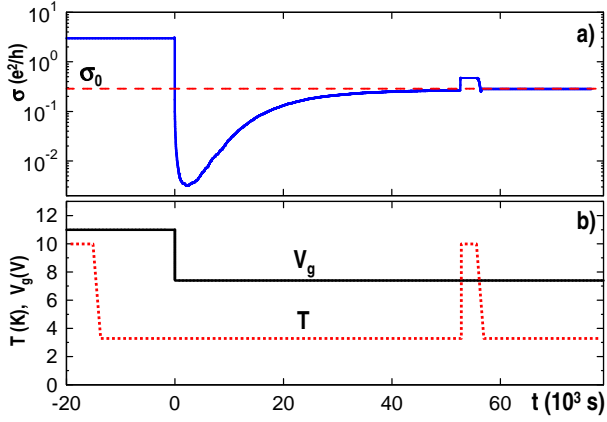,width=8cm,clip=}} \caption{(Color online) (a) $\sigma(t)$ for
$V_{g}^{i}=11$~V [$n_s(10^{11}$cm$^{-2})=20.26$], $V_{g}^{f}=7.4$~V [$n_s(10^{11}$cm$^{-2})=4.74$], and
$T=3.3$~K. (b) Experimental protocol: $V_g(t)$ and $T(t)$. \label{fig:example1}}
\end{figure}

The experiment consists of the following procedure.  The sample is cooled from 10~K to the measurement
temperature $T$ with an initial gate voltage $V_{g}^{i}$.  Then, at $t=0$, the gate voltage is switched rapidly
(within 1~s) to a final value $V_{g}^{f}$, and $\sigma(t,V_{g}^{f},T)$ measured.  In general, the results did
not depend on any of the following: initial temperature, as long as it was $\geq 10$~K; $V_{g}^{i}$; the cooling
time, which was varied between 30 minutes and 10 hours;  the time the sample was kept at 10~K (from 5 minutes to
8 hours); the time the sample spent at the measurement $T$ before $V_g$ was changed (from 5 minutes to 8 hours).
Figure~\ref{fig:example1} shows a typical
%
%
experimental run with $\sigma(t>0)$ exhibiting a rapid ($<10$~s) initial drop followed by a slower
relaxation~\cite{RC}. After going through a minimum, $\sigma(t)$ increases and approaches a value
$\sigma_0(V_{g}^{f},T)$. A subsequent warm-up to $10$~K and a cool down to the same measurement $T=3.3$~K, while
keeping the gate voltage fixed at $V_{g}^{f}$, shows that $\sigma_0(V_{g}^{f},T)$ represents the equilibrium
conductivity corresponding to the given $V_{g}^{f}$ and $T$.  It is interesting that, even though initially it
drops to a value close to $\sigma_0$, $\sigma$ first goes \textit{away} from equilibrium before it starts
approaching $\sigma_0$ again.

\begin{figure}
\centerline{\epsfig{file=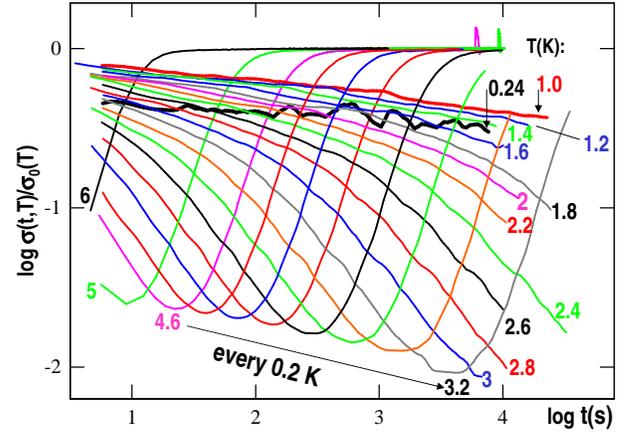,width=8cm,clip=}} \caption{(Color online) $\sigma(t)$ for
$V_{g}^{i}=11$~V, $V_{g}^{f}=7.4$~V, and several $T$, as shown.  For $3.2\leq T($K$)\leq 4.2$, $T$ was changed
in steps of 0.1~K but some traces have been omitted for clarity. Similarly, the data obtained at $0.24<T(K)<$~1
are shown in Fig.~\ref{fig:lowTscaling}(b). \label{fig:allT}}
\end{figure}

At the end of the run, the sample is warmed up to 10~K, gate voltage changed back to the same $V_{g}^{i}$, and
the experiment is repeated at a different $T$ for the same $V_{g}^{f}$. Figure~\ref{fig:allT}
%
%
shows the relaxations of conductivity, $\sigma(t,V_{g}^{f},T)$, normalized to the corresponding
$\sigma_0(V_{g}^{f},T)$ for different $T$.  The minimum in $\sigma$ shifts to longer times with decreasing $T$
until, at sufficiently low $T$, it falls out of the time window of the measurements.  The relaxations obviously
become slower with decreasing $T$ but the shape of the relaxation curves remains similar, suggesting that they
might be collapsed onto a single scaling function.

\begin{figure}
\centerline{\epsfig{file=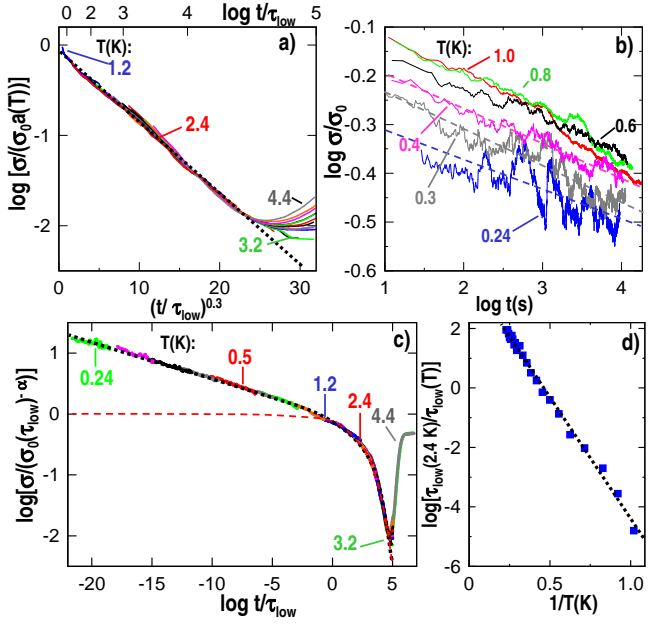,width=8.5cm,clip=}} \caption{(Color online) (a) Scaling of the
Fig.~\ref{fig:allT} data for $t$ just below the minimum in $\sigma(t)$.  The data obey a stretched exponential
dependence on $t$; the dotted line is a fit. The prefactor $a(T)\propto(\tau_{low})^{-\alpha}$,
$\alpha\approx0.07$. (b) At short $t$, relaxations follow a power law $\sigma(t,T)/\sigma_0(T)\propto
t^{-\alpha}$~\cite{logvspower}. The dashed lines are fits with slopes $\alpha=0.07\pm 0.01$.  (c) While the
lowest $T$ data clearly deviate from the stretched exponential dependence (dashed line), all the data are
consistent with the Ogielski relaxation, $\sigma(t,T)/\sigma_{0}(T)\propto
(\tau_{low})^{-\alpha}(t/\tau_{low})^{-\alpha}\exp[-(t/\tau_{low})^{-\beta}]$ (dotted line). The data have been
collapsed with respect to the 2.4~K curve.  (d) Scaling parameter $\tau_{low}$ as a function of $T$.  The dashed
line is a fit with a slope equal to an activation energy $E_a\approx 19$~K; $\tau_{low}(2.4$~K$)\sim 500$~s.
 \label{fig:lowTscaling}}
\end{figure}

First we focus on times just before the minimum in $\sigma(t)$. Figure~\ref{fig:lowTscaling}(a) shows that the
data can be described by a scaling form $\sigma(t,T)/\sigma_0(T)=a(T)g(t/\tau_{low}(T))$, such that
$a(T)\propto(\tau_{low})^{-\alpha}$ ($\alpha\approx 0.07$) and
$g(t/\tau_{low})\sim\exp[-(t/\tau_{low})^{\beta}]$, $\beta=0.3$. We note that this stretched exponential
dependence spans more than 4 orders of magnitude in $t/\tau_{low}$, and 2 orders in $\sigma/\sigma_0$.  At
shorter times, most easily observable at low $T$ within our experimental time window, the stretched exponential
function crosses over to a slower, power-law dependence $\sigma(t,T)/\sigma_0(T)\propto t^{-\alpha}$
[Fig.~\ref{fig:lowTscaling}(b)]~\cite{logvspower}.  In this regime, of course, it is also possible to collapse
the data but, obviously, not in an unambiguous way.  Both power-law and stretched exponential relaxations are
considered to be typical signatures of glassy behavior, and reflect the existence of a broad distribution of
relaxation times.  In spin glasses, for example, the so-called Ogielski function~\cite{Ogielski}, $g(t)\propto
t^{-\alpha}\exp[-(t/\tau(T))^{\beta}]$, describes the relaxations over a wide range of $t$ and $T$ above
$T_g$~\cite{Pappas}.  Our results [Figs.~\ref{fig:lowTscaling}(a), (b)] suggest
%
%
that the Ogielski function might be a good candidate to describe also our data over the entire $t$ and $T$
range. Indeed, Fig.~\ref{fig:lowTscaling}(c) shows that all the data are consistent with the Ogielski relaxation
$\sigma(t,T)/\sigma_{0}(T)\propto (\tau_{low})^{-\alpha}(t/\tau_{low})^{-\alpha}\exp[-(t/\tau_{low})^{-\beta}]$
over about 25 orders of magnitude in $t/\tau_{low}$.  Since it was impractical to perform measurements below 1~K
long enough to obtain a substantial overlap between different $T$ curves, we limit our analysis of
$\tau_{low}(T)$ to the $T\geq 1$~K range, where the data collapse well.  Even within this relatively narrow
range of $T$, the scaling parameter $\tau_{low}$ varies over 7 orders of magnitude
[Fig.~\ref{fig:lowTscaling}(d)], and the Arrhenius function, $1/\tau_{low}=k_{0}\exp(-E_{a}/T)$ ($k_0\sim
10$~s$^{-1}$, $E_{a}\approx 19$~K for $V_{g}^{f}=7.4$~V), provides an excellent fit to the
data~\cite{znucomment}.

\begin{figure}
\centerline{\epsfig{file=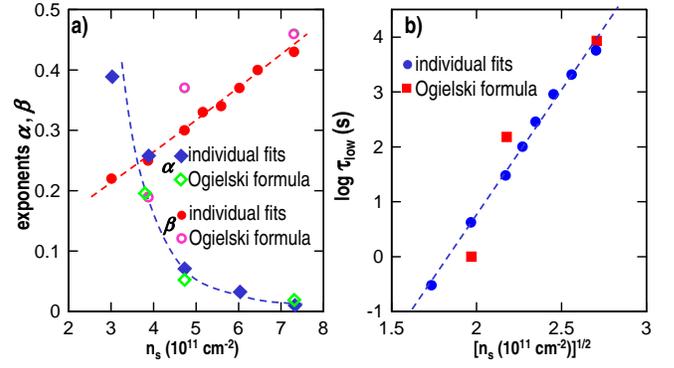,width=8.5cm,clip=}} \caption{(Color online) (a) The exponents $\alpha$
and $\beta$ \textit{vs.} $n_s$.  Dashed lines guide the eye. (b) $\log\tau_{low}$ plotted \textit{vs.}
$n_{s}^{1/2}$. The values of $\tau_{low}$ correspond to $T=3$~K. The dashed line is a fit with a slope $(\log
e)\gamma\approx4.5 (10^{-11}$cm$^{2})^{1/2}$. \label{fig:exponents}}
\end{figure}

The experiment was repeated for different values of $V_{g}^{f}$, allowing us to map out the density dependence
of the exponents $\alpha$ and $\beta$, and of the parameter $\tau_{low}$.  The results were obtained either by
fitting the data to the appropriate functional form at a fixed $T$ [as in Figs.~\ref{fig:lowTscaling}(a) and
(b)] or from the scaling of the different $T$ data and the fit to the Ogielski function [as in
Fig.~\ref{fig:lowTscaling}(c)].  Figure~\ref{fig:exponents} shows that the two methods yield very similar
%
%
results.  In particular, we find that $\alpha$ decreases with $n_s$, and vanishes at
$n_s(10^{11}$cm$^{-2})\approx 7.5\approx n_g$, \textit{i.~e.} exactly where the onset of glassy dynamics was
found to take place from the noise measurements~\cite{SBPRL}.  At the same time, the exponent $\beta$ grows with
increasing $n_s$, indicating that the relaxations become faster, as expected.  For all $n_s$ in this range,
$1/\tau_{low}=k_0(n_s)\exp(-E_a/T)$ with $E_a\approx 20$~K independent of $n_s$, but the prefactor has a very
strong density dependence, $k_0(n_s)\propto\exp (-\gamma n_{s}^{1/2})$ [Fig.~\ref{fig:exponents}(b)]. Since the
ratio of the Fermi energy to Coulomb energy in 2D systems $1/r_s=E_F/U\propto
n_{s}^{1/2}$~\cite{2DMIT-reviews,EFcomment}, our result strongly suggests that Coulomb interactions between 2D
electrons play a dominant role in the observed slow dynamics.  As $n_s$ is increased further, beyond $n_g$,
there are still some visible relaxations, albeit with an amplitude $\sigma/\sigma_0$ that is too small to make
reliable fits.  In fact, the amplitude of the relaxations decreases with $n_s$, and vanishes when
$n_s(10^{11}$cm$^{-2})\sim 30$, where $\sigma_0\sim 3$e$^2/h$, \textit{i.~e.} $k_{F}l\sim1$. Therefore, all the
phenomena discussed in this paper take place in the regime of strong disorder.  We recall that, in general,
electron-electron interactions are enhanced in the presence of disorder~\cite{ES,LR}.

\begin{figure}
\centerline{\epsfig{file=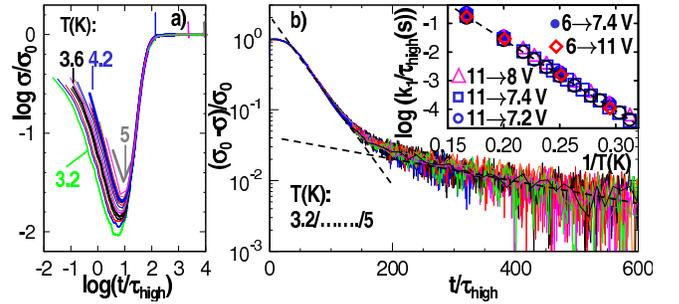,width=8.5cm,clip=}} \caption{(Color online) (a) Scaling of the
Fig.~\ref{fig:allT} data at long $t$, \textit{i.~e.} above the minimum in $\sigma(t)$.  The data have been
collapsed with respect to the 5~K curve. (b)  The scaling function, which describes the approach to equilibrium
value $\sigma_0$ at long times.  The dashed lines are fits with slopes $k_1\approx2.89\times 10^{-2}$ and
$k_2\approx4.37\times 10^{-3}$ for the faster and slower exponential processes, respectively.  (The
corresponding relaxation times are $\tau_{high}/k_1$ and $\tau_{high}/k_2$.) Inset: $T$-dependence of
$\tau_{high}/k_1$ for different $V_{g}^{i}$ and $V_{g}^{f}$ as shown on the plot.
 The dashed line is a fit with the slope equal to
 an activation energy $E_A\approx 57$~K.\label{fig:highTscaling}}
\end{figure}

Finally, at times above the minimum in $\sigma(t)$ (Fig.~\ref{fig:allT}), relaxations at different $T$ approach
the corresponding equilibrium values $\sigma_0(T)$. Here all the data can be described by a scaling form
$\sigma(t,T)/\sigma_0(T)=f(t/\tau_{high}(T))$ [Fig.~\ref{fig:highTscaling}(a)], where
%
%
$f(t/\tau_{high})$ is a simple exponential function with the characteristic relaxation time $\tau_{high}/k_1$.
For some $n_s$ [see Fig.~\ref{fig:highTscaling}(b)], another slower, simple exponential process also seems to
exist with the same $T$-dependence of its characteristic relaxation time $\tau_{high}/k_2$, but its presence
does not seem to depend on $n_s$ in any systematic way.  We find that, unlike $\tau_{low}$, $\tau_{high}/k_1$
(and $\tau_{high}/k_2$, if present) does not depend on $n_s$ in the regime studied.  On the other hand,
$k_1/\tau_{high}=(2.7\times 10^3$~s$^{-1})\exp(-E_A/T)$ with $E_A\approx57$~K independent of $n_s$
(Fig.~\ref{fig:highTscaling}(b) inset).

Several striking features of our data stand out.

(1) At long times, the relaxation is exponential with a characteristic time $\tau_{high}/k_1$ that diverges
exponentially as $T\rightarrow 0$.  This means that, at any finite $T$, the 2DES will reach equilibrium after a
sufficiently long time $t\gg(\tau_{high}/k_1)$.  For example, $\tau_{high}/k_1\sim 30$~s at $T=5$~K but, already
at $T=1$~K, $\tau_{high}/k_1\sim 10^{13}$~years, which exceeds the age of the Universe by several orders of
magnitude.  Therefore, even though, strictly speaking, the system appears to be glassy only at $T=0$, at low
enough $T$ the dynamics is glassy on all experimentally accessible time scales.  It is interesting that the
recent study of the 2D Coulomb glass model has also found~\cite{Grempel} an exponential divergence of the
equilibration time as $T\rightarrow 0$, signaling a transition at $T_g=0$.

(2) In the glassy regime, \textit{i.~e.} on short enough time scales $t<(\tau_{high}/k_1)$, the relaxations are
strongly nonexponential.  They obey the Ogielski form $\sigma(t,T)/\sigma_{0}(T)\propto
t^{-\alpha}\exp[-(t/\tau_{low})^{-\beta}]$ with $0<\alpha<0.4$ and $0.2<\beta<0.45$, where
$\tau_{low}\propto\exp(\gamma n_{s}^{1/2})\exp(E_a/T)$, $E_a\approx 20$~K.  Since $\tau_{low}$ diverges as
$T\rightarrow0$, the relaxations attain a pure power law form $\propto t^{-\alpha}$ at $T=0$.  We note that the
Ogielski form of the relaxation and the divergence of $\tau_{low}$ are consistent with the general scaling
arguments~\cite{scaling} near a continuous phase transition occurring at $T_g=0$.  Similar scaling is observed
in spin glasses above $T_g$~\cite{Pappas}.

The very pronounced and precise dependence of $\tau_{low}$ on $n_s$ strongly suggests that Coulomb interactions
between 2D electrons play a dominant role in the observed out-of-equilibrium dynamics.  In InO$_x$ films, the
most extensively studied example of an electron glass~\cite{films2,films2-Zvi,films2-Zvi2}, the dependence of
the relaxation time on the carrier density has been interpreted as evidence that the slow dynamics is due to
electronic relaxation rather than glassiness of the extrinsic degrees of freedom~\cite{films2-Zvi}.

(3) The 2DES equilibrates only after it first goes farther away from equilibrium.  This overshooting of $\sigma$
manifests itself as either a minimum [Fig.~\ref{fig:example1}(a)] or a maximum in $\sigma(t)$ depending on the
direction of $V_g$ change.  For example, if $V_{g}^{i}<V_{g}^{f}$, $\sigma(t)$ first increases, overshoots
$\sigma_0$ and reaches a maximum, and then decreases exponentially to $\sigma_0$. An analogous overshooting
during relaxation is known to occur in orientational~\cite{over-orient} and spin glasses~\cite{over-spin} but
has not been observed in InO$_x$ films~\cite{films2-Zvi2}.  We note an important difference between our
experiment and those on InO$_x$ films.  In InO$_x$, the change in the carrier density due to variations in $V_g$
is typically $\sim 1$\%, and the system remains deep in the insulating state. In our case, $n_s$ is changed up
to a factor of 7, and the 2DES may go from the conducting to the insulating regime. Recent numerical work has
shown~\cite{over-sim} that such a strong perturbation can lead to nontrivial dynamics and, in particular, to a
``roundabout'' relaxation, where system equilibrates only after it once goes farther away from equilibrium.

In summary, we have studied the relaxations of $\sigma$ in a strongly disordered 2DES after excitation far from
equilibrium by a rapid change of $n_s$ at low $T$.  The data strongly suggest (a) the transition to a glassy
phase as $T\rightarrow0$, and (b) the Coulomb interactions between 2D electrons are primarily responsible for
the observed out-of-equilibrium dynamics.  Further work is needed to identify the microscopic details of the
relaxation phenomena.

We are grateful to I. Rai\v{c}evi\'{c} for technical assistance, and V. Dobrosavljevi\'c for useful discussions.
This work was supported by NSF grant No. DMR-0403491, and NHMFL through NSF Cooperative Agreement No.
DMR-0084173.  D.~P. would like to acknowledge the hospitality of the Aspen Center for Physics.

\newcommand{\noopsort}[1]{} \newcommand{\printfirst}[2]{#1}
  \newcommand{\singleletter}[1]{#1} \newcommand{\switchargs}[2]{#2#1}


\begin{thebibliography}{24}
\bibitem[{RC()}]{RC}
\bibinfo{note}{The RC charging times of the device and the circuit are at most 10~ms}.
\bibitem[{log()}]{logvspower}
\bibinfo{note}{While it is difficult to distinguish between a logarithm and a power-law
for the case shown in Fig.~\ref{fig:lowTscaling}(b), other data indicate that the power law provides a better
fit}.
\bibitem[{znu()}]{znucomment}
\bibinfo{note}{If the values for $\tau_{low}(T<1$~K) that were obtained assuming the Ogielski form
[Fig.~\ref{fig:lowTscaling}(c)] are also included in the fit, then all the data are described well by
$1/\tau_{low}\approx (10^{2}$s$^{-1})\exp(-c/T^b)$, where $b\sim 0.7$ but there is a large uncertainty in the
corresponding activation energy. Similar sub-Arrhenius $T$-dependence of the relaxation time has been found in
some theoretical models that display glassy dynamics~\cite{subArrhenius-model}, and also observed in some glassy
thin films~\cite{subArrhenius-exp}. We note that none of the main conclusions of our work depend on the exact
value of the exponent $b$}.
\end{thebibliography}
\end{document}